\documentclass[12pt]{article}
\usepackage{amsmath,amssymb,epsfig}
\begin{document}
\parindent=0pt
\parskip=6pt
\rm

\vspace*{1cm}
\begin{center}
{\bf ABOUT THE MAGNETIC FLUCTUATION EFFECT ON THE PHASE
TRANSITION TO SUPERCONDUCTING STATE IN Al}

\vspace{0.5cm}

  D. V. SHOPOVA$^{\ast}$,
 T. P. TODOROV$^{\dag}$, T. E. TSVETKOV and D. I. UZUNOV$^{\ddag}$

{\em  CPCM Laboratory, G. Nadjakov Institute of Solid
 State Physics,\\
 Bulgarian Academy of Sciences, BG-1784 Sofia, Bulgaria.} \\
\end{center}

  $^{\ast}$ Corresponding author.

 $^{\dag}$ Permanent address: Joint Technical College at the Technical
University of Sofia.

$^{\ddag}$ Temporal address: MPI-PKS, N\"{o}thnitzer Str. 38,
 01187 Dresden, Germany.

\vspace{1cm}


\vspace{0.5cm}


\vspace{0.3cm}

\begin{abstract}

The free energy and the order parameter profile near the phase transition to
the superconducting state in bulk Al samples are calculated within a
mean-field-like approximation. The results are compared with those for thin
films.

\end{abstract}

{\bf 1. Introduction}

In this letter we discuss in details the fluctuation-induced weakly-first order
phase transition in type I superconductors known as Halperin-Lubensky-Ma (HLM)
effect~\cite{HLM:1974}. Our numerical results for the free energy and the order
parameter profile are presented for Al which is the
best substance for an experimental observation of the effect. Three dimensional
(3D), i.e., bulk Al samples are considered. The results are
compared with those for quasi-2D (two dimensional) Al films~\cite{FSU:2001}.
The possibility for an experimental observation of the effect in Al is
briefly discussed. The paper is intended to establish by a quantitatively
precise evaluation of measurable physical quantities, the difference in the
magnitude of the effect in three dimensional (3D) and quasi-2D samples. It
seems interesting to justify the experimental search of the effect in suitable
films of type I superconductors where the same effect is relatively strong and
can be observed experimentally as predicted in Ref.~\cite{FSU:2001}. This task
includes an entire investigation of the quasi-2D and 3D cases as well as a
detailed description of the difference between them. Here we shall
 compare our results with those in Refs.~\cite{HLM:1974, FSU:2001}.

{\bf 2. Model and Results}

Let us remember that in 3D type I superconductors the magnetic fluctuations
$\delta {\vec{H}} = \nabla
\times \delta\vec{A}$ at the critical point $T_{c0}$ corresponding to zero mean
magnetic field $\vec{H}_0 = (\vec{H} - \delta\vec{H})$~\cite{LP:1980,
UZ:1993} produce
a very small latent heat evaluated in Ref.~\cite{HLM:1974} and a jump of the
superconducting
order parameter $\psi$ which is calculated for the first time in the present
report; $\delta\vec{A} = (\vec{A} - \vec{A}_0)$ is the fluctuation part of the
vector potential $\vec{A}$ of the magnetic field $\vec{H}$, whereas $\vec{A}_0$
corresponds to the mean value $\vec{H}_0$. The same type of fluctuation-induced
first-order phase transition is
predicted in the scalar electrodynamics, early universe theories, and in liquid
crystals. Moreover, it is generally believed that the same transition
should occur in any system, described by a gauge invariant interaction
between a scalar field such as the superconducting order parameter
$\psi(\vec{x})$ and a vector gauge field as the vector potential
$\vec{A}(\vec{x})$ of the magnetic field $\vec{H}(\vec{x}) = \nabla \times
\vec{A}(\vec{x})$ in superconductors, provided the characteristic lengths in
the system satisfy certain conditions; see, e.g.,
Refs.~\cite{FSU:2001, UZ:1993, FH:1999}. These notes justify the significance
of the effect and the importance of its investigation.

 For our aims we shall use
the ``mean-field-like approximation" explained in Refs.~\cite{HLM:1974,
FSU:2001, UZ:1993}; the renormalization group treatment of the effect has been
recently reviewed in Ref.~\cite{FH:1999}. Using the notations from
Ref.~\cite{LP:1980}
the Ginzburg-Landau free energy of a superconductor can be written in the form
\begin{equation}
F = \int d^3 x \left[ a|\psi|^2 + \frac{b}{2}|\psi|^4 +
\frac{\hbar^2}{4m}\left|\left(\nabla - \frac{2ie}{\hbar c}\vec{A}\right)\psi
\right|^2 + \frac{1}{16\pi}\sum_{i,j=1}^{3} \left( \frac{\partial
A_i}{\partial x_j} - \frac{\partial A_j}{\partial x_i} \right)^2 \right]\;.
\end{equation}
Within the mean-field-like approximation~\cite{HLM:1974}, the fluctuations of
the superconducting order parameter are neglected. Having in mind that in
type I superconductors the only stable ordered phase is the Meissner phase, we
shall consider the equilibrium order parameter as spatially independent:
$\psi(\vec{x}) \approx \psi_0$; other details of this approximation and its
limitations are given in Refs.~\cite{FSU:2001, UZ:1993}. For $\vec{H}_0=0$ we
set $\vec{A}_0 = 0$ and, therefore, $\vec{A} \equiv \delta\vec{A}$.

 The integration of the vector potential fluctuations $\delta\vec{A}(\vec{x})$
in the partition function is made with the help of a loop-like
expansion~\cite{FSU:2001}. The next step is the accomplishment of
the Landau expansion of the free energy in power series of the magnitude
$|\psi_0|$ of the
order parameter $\psi_0$. In this way, within the Landau expansion to order
$|\psi|^6$ we obtain the effective free
energy $f_{\mbox{\footnotesize{eff}}} = F_{\mbox{\footnotesize{eff}}}/V$, where
V is the volume of the superconductor, and \begin{equation}
f_{\mbox{\footnotesize{eff}}} = \tilde{a}|\psi|^2 + \frac{\tilde{b}}{2}|\psi|^4
+ q|\psi|^3 + c|\psi|^6\;.
 \end{equation}
The above expression contains new Landau parameters,
\begin{equation}
\tilde{a} = a + \frac{\rho_0k_BT\Lambda}{2\pi^2},\;\;\;\;\;\;
\tilde{b} = b + \frac{\rho_0^2k_BT}{2\pi^2\Lambda}\;,
 \end{equation}
renormalized by the fluctuation effects, and additional new parameters
\begin{equation}
q = - \frac{\rho_0^{3/2}k_BT}{6\pi},\;\;\;\;\;\;
c = - \frac{\rho_0^3k_BT}{18\pi^2\Lambda^3}\;,
 \end{equation}
which are also generated by the magnetic field fluctuations. In Eqs.~(3) and
(4), $\rho_0 = (8\pi e^2/mc^2)$, and $\Lambda = (\pi/\xi_0)$, where $\xi_0 =
(\hbar^2/4m\alpha_0T_{c0})$ is
the so-called zero-temperature coherence length~\cite{LP:1980}; $\alpha_0$ is
related to the Landau parameter $a$ through $a = \alpha_0T_{c0}t$.Here $t =
(T-T_{c0})/T_{c0}$ is the reduced temperature distance from the critical
 temperature $T_{c0}$.  Note, that the $|\psi|^6$ term is derived for the
first time. Moreover, the $\rho_0-$term in $\tilde{b}$, see Eqs.~(3), which
was neglected in Ref.~\cite{HLM:1974} (as mentioned for the first time in
Ref.~\cite{CLN:1978}), is also calculated for the first time in the present
investigation. As we shall see, the $\rho_0-$ contribution to $\tilde{b}$ is
small ($\sim 0.1b$) for 3D-systems, but it becomes relatively bigger ($\sim b$)
in quasi-2D
films~\cite{FSU:2001} and generally cannot be omitted. In all other respects
the Eqs. (2)~-~(4) are consistent with the results in the preceding
papers~\cite{HLM:1974, CLN:1978}.

The $|\psi|^3-$term in Eq.~(2) describes a first order transition
(see, e.g., Ref.~\cite{UZ:1993}). The theory shows~\cite{HLM:1974}
that this fluctuation-induced first order transition should be
well established in type I superconductors with a small
Ginzburg-Landau parameter $\kappa = \lambda_0/\xi_0$, where
$\lambda_0$ is the so-called zero-temperature London penetration
depth~\cite{LP:1980, UZ:1993}. The Al is the type I superconductor
with a minimal parameter $\kappa \sim 10^{-2}$ and therefore, this
substance is convenient for a discussion of the HLM effect, as
mentioned for the first time in Ref.~\cite{HLM:1974}. Using the
experimental values for the critical temperature $T_{c0} = 1.19$K,
the zero-temperature coherence length $\xi_0 = 1.6\times 10^{-4}$
cm, and the zero-temperature critical magnetic field $H_c(0) = 99$
G, as well as applying certain relations between $\alpha_0$ and
$\xi_0$, and between $b$ and $H_c(0)$ (see Ref.~\cite{LP:1980}),
it is easy to calculate the effective free energy for Al:
\begin{eqnarray}
f_{\mbox{\footnotesize{eff}}}(\varphi) & = & 389.21 \{ 2 \left[ t + 0.972
\times 10^{-4}(1+t)\right] \varphi^2 + 1.117(1+t)\varphi^4 \\ \nonumber
&& - 0.7053\times 10^{-2} (1+t)\varphi^3 -
31.1(1+t) \varphi^6 \}\;,
\end{eqnarray}
where $\varphi = |\psi_0|/|\psi_{00}|$ is the dimensionless order parameter;
$|\psi_{00}|$ denotes $|\psi_0|$ at $T=0$. The contribution of the
$\varphi^6-$term is very small for 3D Al samples and we shall neglect this term
in our further discussion.

\begin{figure}
\begin{center}
\epsfig{file=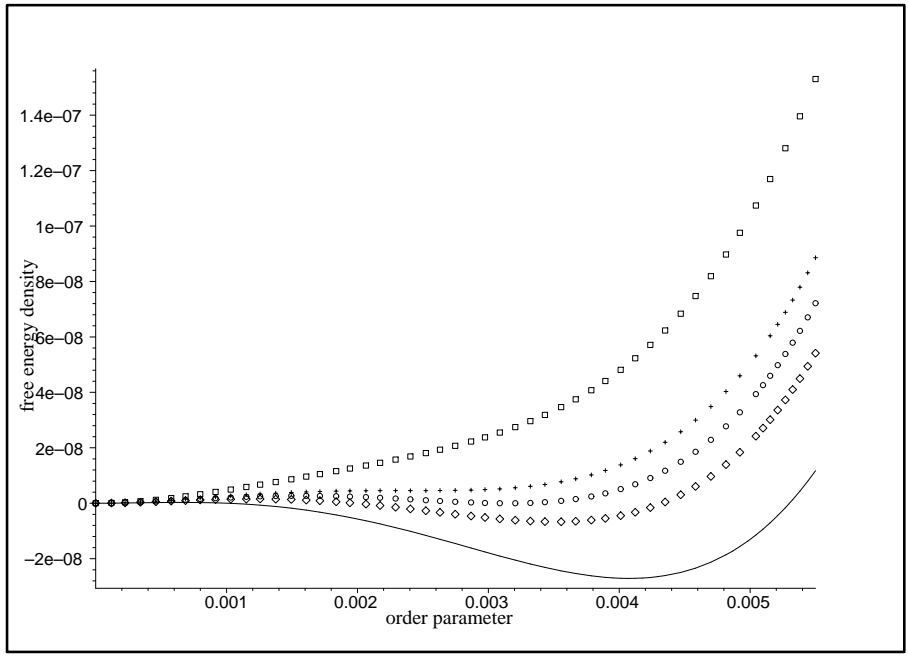,angle=-90, width=12cm}\\
\end{center}
\caption{Curves representing the free energy function (5) for
$c\equiv 0$ and five values of $t$: $t= 9\times
10^{-6}\;(\square)$,$\; t= 6.263\times 10^{-6}\;(+)$, $t = 5.567
\times 10^{-6}\;(\circ)$, $t = 4.800\times10^{-6}\;(\diamond)$, $t
= 3.000 \times 10^{-6}(-)$.}
 \label{d1f1.fig}
\end{figure}

The free energy (5) is shown
in Fig.~1 for five values of $t$, and for $c\varphi^6 \approx 0$. The Fig.~1
gives for the first time a graphical image of the weakly-first order phase
transition predicted in Ref.~\cite{HLM:1974} for 3D superconductors of type I.
All curves have a trivial minimum at $\varphi = 0$ which describes the normal
state. The positive minima of the free energy $f_{\mbox{\footnotesize
eff}}(\varphi)$,  $f^{\mbox{\footnotesize min}}_{\mbox{\footnotesize eff}} >
0$, corresponding to $\varphi > 0$ describe the superconducting
Meissner phase which is metastable for the respective temperatures. The
negative minima, $f^{\mbox{\footnotesize min}}_{\mbox{\footnotesize eff}} < 0$,
 corresponding to $\varphi > 0$ represent the free
energy of stable superconducting states and occur in certain narrow temperature
interval, approximately evaluated for the first time in Ref.~\cite{HLM:1974}.
The curve marked with circles ($\circ$) exhibits a minimum
$f_{\mbox{{\footnotesize eff}}}^{\mbox{{\footnotesize min}}}(\varphi=0.00316)
= 0$ which is equal to the free energy of the normal phase.
Therefore, this minimum corresponds to the equilibrium phase transition
temperature.

\begin{figure}
\begin{center}
\epsfig{file=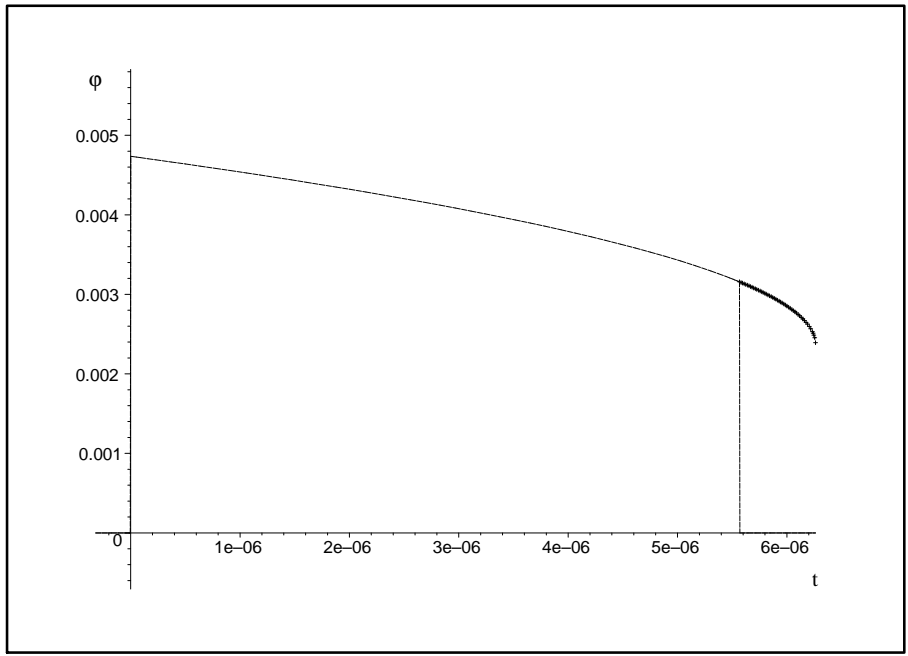,angle=-90, width=12cm}\\
\end{center}
\caption{Order parameter profile near $T_{c0}$. The vertical line
at $t = 5.567\times 10^{-6}$ indicates the equilibrium jump of the
order parameter.} \label{d1f2.fig}
\end{figure}

The standard analysis~\cite{UZ:1993} of the equations
$f_{\mbox{\footnotesize eff}}(\varphi) = 0$
 and $(\partial f_{\mbox{\footnotesize eff}}/\partial \varphi) = 0$ gives
the properties of the
order parameter $\varphi$. This important quantity is depicted in Fig.~2 as a
function of the reduced temperature $t$. Fig.~2 shows that
the stable superconducting states, corresponding to a negative
effective free energy occur for temperatures defined by $0 < t<
t_{\mbox{{\footnotesize eq}}}=5.567\times 10^{-6}$,
i.e. for $T_{c0} < T <
T_{\mbox{\footnotesize{eq}}} =  (1 + 5.567\times 10^{-6})T_{c0}$. At the
equilibrium transition temperature ($T=
T_{\mbox{\footnotesize{eq}}}$) the normal and the superconducting phases are
equally stable, and from $T > T_{\mbox{\footnotesize{eq}}}$ up to the
temperature $T^{\ast} = (1 + 6.262 \times 10^{-6})T_{c0}$ defined by $t^{\ast}
= 6.262\times 10^{-6}$, the superconducting phase is metastable. Above
$T^{\ast}$ the equation $f_{{\mbox{\footnotesize eff}}}(\varphi) = 0$ has no
solutions of type $\varphi > 0$, and therefore the 3D Al does not possess any
superconducting states. At the equilibrium phase transition point
$T_{\mbox{\footnotesize{eq}}}$ the order parameter
 $\varphi$ undergoes an equilibrium jump from the value $\varphi =
0.00316$ to zero, provided quite special circumstances do not ensure an
overheating of the superconducting states. The latter possibility is usual for
first order transitions, where metastable states above the equilibrium phase
transition temperature are possible.

    The calculated value of the equilibrium jump of the order parameter is very
small for an experimental observation by transport experiments. This is so
because of  the weakness of the HLM effect for 3D. As pointed out in a recent
study~\cite{FSU:2001, STU:2002}, the
same effect is much better pronounced in quasi-2D Al and, therefore, transport
experiments in such films could be successful.

{\bf 3. Conclusion}

For a better comparison between the 3D and quasi-2D Al superconductors, let us
note that the order parameter jumps for thin Al films of thicknesses 0.1$\mu$m
and 1$\mu$m  are $\varphi = 0.032$ and $\varphi = 0.013$,
respectively~\cite{STU:2002}. Besides, the temperature differences
$(T_{\mbox{\footnotesize eq}} - T_{c0})$ and $(T^{\ast} - T_{c0})$ are about
$10^3$ times bigger than the respective differences for 3D Al discussed above.
These results imply that the expected specific heat capacity and latent heat of
the first order phase transition in quasi-2D Al films are much bigger than the
respective quantities in 3D Al samples. So, the thermodynamic
experiments done with Al films may prove the HLM as well.


\newpage
\vspace*{1cm}


\begin{thebibliography}{9}
\bibitem{HLM:1974}
B. I. Halperin, T. C. Lubensky, and S. K. Ma,
{\em Phys. Rev. Lett.} {\bf 32}, 292 (1974).

\bibitem{FSU:2001}
R. Folk, D. V. Shopova and D. I. Uzunov, {\em Phys. Lett.} {\bf A 281}, 197
(2001).

\bibitem{LP:1980}
 E. M. Lifshitz and L. P. Pitaevskii, {\em Statistical Physics}, Part 2,
[Landau and Lifshitz Course of Theoretical Physics, vol. 9]
 (Pergamon Press, Oxford, 1980).

\bibitem{UZ:1993}
 D. I. Uzunov, {\em Theory of Critical
 Phenomena}
(World Scientific, Singapore, 1993).


\bibitem{FH:1999}
 R. Folk and Yu. Holovatch, in: {\em Correlations, Coherence, and Order},
 ed. by D. V. Shopova and D. I. Uzunov
 (Kluwer Academic/Plenum Publishers, New York-London, 1999), p. 83.

\bibitem{CLN:1978}
 J-H. Chen, T. C. Lubensky, and D. R. Nelson, {\em Phys. Rev.} {\bf B17}, 4274
(1978).

\bibitem{STU:2002}
D. V. Shopova, T. P. Todorov, and D. I. Uzunov, unpublished (2002).

\end{thebibliography}
\end{document}